\documentclass[sigconf]{acmart}

\copyrightyear{2024}
\acmYear{2024}
\setcopyright{rightsretained}
\acmConference[SIGIR '24]{Proceedings of the 47th International ACM SIGIR Conference on Research and Development in Information Retrieval}{July 14--18, 2024}{Washington, DC, USA}
\acmBooktitle{Proceedings of the 47th International ACM SIGIR Conference on Research and Development in Information Retrieval (SIGIR '24), July 14--18, 2024, Washington, DC, USA}
\acmDOI{10.1145/3626772.3657765}
\acmISBN{979-8-4007-0431-4/24/07}

\makeatletter
\gdef\@copyrightpermission{
 \begin{minipage}{0.7\columnwidth}
  \href{https://creativecommons.org/licenses/by/4.0/}{This work is licensed under a Creative Commons Attribution International 4.0 License.}
 \end{minipage}
 \vspace{5pt}
}
\makeatother

\usepackage{graphicx}
\usepackage{subfigure} 
\usepackage{todonotes}
\usepackage[normalem]{ulem}
\usepackage{color,soul}
\usepackage{xcolor}
\usepackage[para]{footmisc}
\definecolor{mygray}{HTML}{999999}

\newcommand{\red}[1]{\textit{\textcolor{mygray}{#1}}} %

\newcommand{\crc}[1]{#1}

\renewcommand{\mid}{\hspace{0.2em}|\hspace{0.2em}}

\begin{document}

\title{Neural Passage Quality Estimation for Static Pruning}

\author{Xuejun Chang}
\orcid{0009-0006-6531-2840}
\affiliation{%
  \institution{University of Glasgow}
  \city{Glasgow}
  \country{United Kingdom}
}
\email{x.chang.2@research.gla.ac.uk}

\author{Debabrata Mishra}
\orcid{0009-0003-8483-3738}
\affiliation{%
  \institution{University of Glasgow}
  \city{Glasgow}
  \country{United Kingdom}
}
\email{2348074M@student.gla.ac.uk}

\author{Craig Macdonald}
\orcid{0000-0003-3143-279X}
\affiliation{%
  \institution{University of Glasgow}
  \city{Glasgow}
  \country{United Kingdom}
}
\email{{first}.{last}@glasgow.ac.uk}

\author{Sean MacAvaney}
\orcid{0000-0002-8914-2659}
\affiliation{%
  \institution{University of Glasgow}
  \city{Glasgow}
  \country{United Kingdom}
}
\email{{first}.{last}@glasgow.ac.uk}

\begin{abstract}
Neural networks---especially those that use large, pre-trained language models---have improved search engines in various ways. Most prominently, they can estimate the relevance of a passage or document to a user's query. In this work, we depart from this direction by exploring whether neural networks can effectively predict which of a document's passages are unlikely to be relevant to \textit{any} query \crc{submitted to the search engine}. We refer to this query-agnostic estimation of passage relevance as a passage's \textit{quality}. \crc{We find that our novel methods for estimating passage quality allow passage corpora to be pruned considerably while maintaining statistically equivalent effectiveness; our best methods can consistently prune $>$25\% of passages in a corpora, across various retrieval pipelines. Such substantial pruning reduces the operating costs of neural search engines in terms of computing resources, power usage, and carbon footprint---both when processing queries (thanks to a smaller index size) and when indexing (lightweight models can prune low-quality passages prior to the costly dense or learned sparse encoding step).} This work sets the stage for developing more advanced neural ``learning-what-to-index'' methods.

\vspace{1.5em}
\hspace{1.8em}\includegraphics[width=1.25em,height=1.25em]{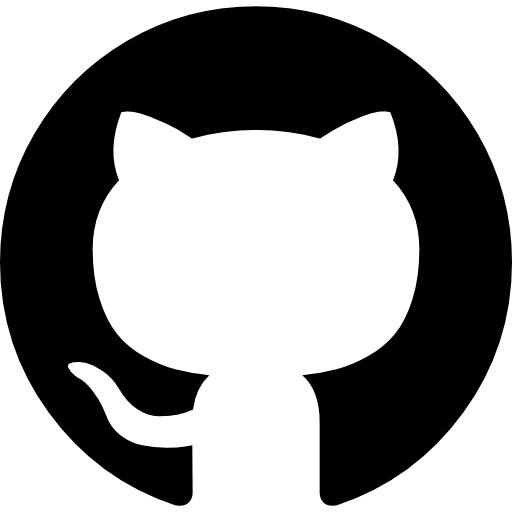}\hspace{.3em}
\parbox[c]{\columnwidth}
{
    \vspace{-.55em}
    \href{https://github.com/terrierteam/pyterrier-quality}{\nolinkurl{https://github.com/terrierteam/pyterrier-quality}}
}
\vspace{-0.5em}
\end{abstract}



\maketitle

\vspace{1em}
\section{Introduction}

Neural language models have markedly improved search engines, especially for estimating the relevance between a query and document~\cite{lin2022pretrained}. Numerous classes of neural models have been proposed to perform relevance estimation---including cross-encoders (e.g., MonoELECTRA~\cite{DBLP:conf/ecir/PradeepLZLYL22}), dense retrieval models (e.g., TAS-B~\cite{tasb}), and learned sparse models (e.g., SPLADE~\cite{formal2021splade})---each with their own trade-offs.
Common across all these approaches is the assumption of a fixed corpus: the approaches define how to score, index, and/or retrieve a given set of documents, but not which documents are worth the processing and storage costs.

Due to efficiency considerations and limitations of their architectures, most neural approaches are designed to work over text of a limited length, typically around a paragraph. Most approaches will either assume that the documents in the corpus are already of limited length (truncating anything that exceeds the maximum) or split the corpus into passages of suitable length using approaches like applying a sliding text window~\cite{DBLP:conf/sigir/HearstP93}, leveraging the document structure~\cite{DBLP:conf/sigir/Callan94}, or applying other (often proprietary) content-based heuristics~\cite{bajaj2016ms}. However, after performing passage segmentation, some of a document's passages are not useful in satisfying \textit{any} information need that users are likely to submit to the engine. Take, for instance, a document from MSMARCO v2~\cite{DBLP:conf/trec/Craswell0YCL21} presented in Figure~\ref{fig:example}. Here, we posit that two of the document's five passages do not contain information that would satisfy any question-answering-style query. We argue that these passages---which we refer to going forward as \textit{low-quality} passages---are harmful to a search engine: they use additional computational and storage resources for content that will likely not be useful to any user. One solution is to apply a tiered indexing approach~\cite{DBLP:conf/la-web/RisvikAL03}, wherein passages that are retrieved frequently are stored in a primary tier, and those retrieved less frequently are stored in lower tiers. Although this addresses retrieval latency overheads, it does not address storage or indexing costs and increases the engine's complexity.

\begin{figure*}
\scalebox{0.95}{
\fbox{
\parbox{\textwidth}{
\small

\textcolor{gray}{(Passage 1, \textit{more valuable})} American/ USA Made 1993 Fender Stratocaster Electric Guitar Tobacco SunBurst with Rosewood Fretboard and Original Hardshell Case. American/ USA Made. 1993 Fender Stratocaster Electric Guitar. Tobacco SunBurst with Rosewood Fretboard and. Original Hardshell Case. SLEEK AND GORGEOUS! Up for sale is my first 'real' guitar.

\vspace{0.05em}

\textcolor{gray}{(Passage 2, \textit{less valuable})} We're not here to talk about that guitar. WE ARE HERE TO CHECK OUT THIS 1993 FENDER STRATOCASTER USA!! Tobacco-SunBurst with a rosewood fretboard, I purchased it new in Seattle late 1993 and have been the only owner/ player.

\vspace{0.05em}

\textcolor{gray}{(Passage 3, \textit{less valuable})} As you can see in the photos it is still in very nice condition being a guitar that has been played...and loved. Its of course not without a couple of bumps and bruises; all of which are quite simply cosmetic and none of which affect the integrity of this fine instrument.

\vspace{0.05em}

\textcolor{gray}{(Passage 4, \textit{more valuable})} The original white pickguard, pickup covers and knobs have been replaced with black..which I feel really helps the SunBurst stand-out beautifully and gives it a much cleaner look complimenting the rosewood fretboard. The original white pickguard will be included in the sale.

\vspace{0.05em}

\textcolor{gray}{(Passage 5, \textit{more valuable})} Frets are good and have a lots of life. As with any collectible/ used guitar we sell and/or buy, we advise strong consideration for having it tuned and setup to your liking once you receive it. You will be much happier with your investment. Original hardshell case is included as pictured.

}}}
\vspace{-1em}
\caption{\crc{Example from \texttt{msmarco\_doc\_01\_1225433927} showing that not all passages within a document are necessarily valuable.}}
\label{fig:example}
\end{figure*}

Instead, in this work, we ask the question: \textit{Can neural methods identify low-quality passages before indexing?} By \textit{pruning} (i.e., removing) low-quality passages entirely early in the ingestion process, a system can avoid all future indexing, storage, and retrieval costs associated with the passage. We explore several existing lexical methods for estimating a passage's quality and find that although they have been shown to be effective signals in an ensemble scoring method~\cite{ZhouDQ2005}, they do not provide strong enough quality signals to consistently outperform a random \crc{pruning} baseline. We therefore explore a variety of potential neural passage quality estimation techniques, including ones that use unsupervised signals (e.g., a passage's perplexity from a language model), latent signals (e.g., the magnitude of a dense retrieval model's vector), and direct supervision (e.g., a model fine-tuned directly on relevance labels). We find that supervised neural passage quality models provide a consistently strong signal for passage pruning, consistently enabling 25\% or more of passages in the MSMARCO corpus to be pruned with statistically equivalent effectiveness downstream on lexical, dense, learned sparse, and re-ranking pipelines. When using a lightweight supervised neural model (e.g., a 4-layer transformer), the pruning process reduces the total neural indexing time, since fewer passages need to be sent to a more costly passage encoder. Finally, we show that the observations hold when moving to substantially larger corpora and corpora in other domains.

We recognise that this work can be viewed from several perspectives. On the one hand, it can be seen as a progression of \textit{static pruning} techniques~\cite{10.1145/383952.383958,Buttcher2006Document-Centric}. In contrast with prior work on static pruning, which focus on removing individual tokens~\cite{Buttcher2006Document-Centric} or token representations~\cite{DBLP:conf/doceng/AcquaviaMT23} from a passage, we investigate ways to remove passages in their entirety. Alternatively, this work can be viewed as an extension of \textit{retrievability}~\cite{DBLP:conf/ictir/Azzopardi15}, which studies the fairness of retrieval systems from a producer perspective by measuring which documents are less prone to retrieval. While we recognise the potential for a passage pruning approach to yield unfair results from a producer perspective, we argue that there are substantial benefits for the search engine operator and searcher perspectives, namely in reduced computational overheads. Further, this work could be viewed as a progression of neural relevance modeling, which presently focuses on ``Mono'' scoring~\cite{DBLP:journals/corr/abs-1901-04085} (how relevant is a document to a query), Pairwise scoring~\cite{DBLP:journals/corr/abs-2101-05667} (which of these two documents are most relevant to a query), and (recently) Listwise scoring~\cite{DBLP:conf/emnlp/0001YMWRCYR23} (sort this set of documents by most relevant to a query). In contrast with this thread, we remove the dependency on an individual query, instead predicting whether a document is likely relevant to \textit{any} query, similar to how a Mono scorer predicts the relevance of a passage compared to \textit{any} other passage in a corpus. Finally, this work could be seen as a form of web spam filtering~\cite{DBLP:journals/ir/CormackSC11}. However, unlike spam, low-quality passages are not necessarily of malicious intent~\cite{DBLP:journals/computer/GyongiG05} (e.g., the low-quality passages in Figure~\ref{fig:example} are just part of the listing's discourse), and by definition, they are unlikely to satisfy an expected information need. In further contrast with common practices in spam detection, we focus exclusively on the textual content of a passage (ignoring features like the source), and train directly on relevance assessments (rather than spam labels).

In summary, we explore techniques for modeling passage quality (i.e., whether a passage is likely relevant to any query) and explore neural models for performing this task. We find that supervised models can prune a substantial proportion of a passage corpus without negatively affecting a variety of retrieval pipelines, while other passage quality approaches struggle to consistently outperform a random baseline. We find that passage pruning can reduce an engine's storage and retrieval overheads and can even reduce dense or learned sparse indexing costs.

\section{Related Work}\label{sec:related}

It has long been recognised that many documents are not relevant for any query, and that some portions of documents need not be indexed for effective retrieval. While stopword removal is a standard technique deployed for reducing the size of an index, many static pruning techniques were proposed to identify documents that would be unlikely to be retrieved, or to only index their postings that would cause them to be retrieved. For instance, some static pruning methods remove terms from the index~\cite{10.5555/1763653.1763665} ({\em term-centric}), or all postings for a given term with low weight~\cite{10.1145/383952.383958}. On the other hand, {\em document-centric} approaches aimed to remove postings from documents that would be unlikely to contribute to the retrieval of that document~\cite{10.1145/1658377.1658378, Buttcher2006Document-Centric,DBLP:conf/ecir/ThotaC11}. \crc{In addition to applying query- and document-centric pruning techniques over lexical indexes, these categories of static pruning methods have also found application in modern dense~\cite{DBLP:conf/doceng/AcquaviaMT23} and learned sparse~\cite{DBLP:conf/sigir/MackenzieDGC20,DBLP:conf/sigir/LassanceLDCT23} indexes.}

Our work is more similar to a third form of static pruning, concerned with identification of entire documents that need not be indexed. For instance, Zheng \& Cox~\cite{10.1007/978-3-642-00958-7_72} aggregated term-level entropy to compute a global document importance measure  -- this can be considered an early form of perplexity calculation on a unigram language model. Others identified a minimum set of documents to index that could well cover common queries~\cite{10.1007/978-3-030-59212-7_13, 10.1145/1935826.1935923}, while Altvingode et al.~\cite{sengor2012} examined page access statistics to prune documents. 

Moreover, with the rising presence of web spam~\cite{DBLP:journals/computer/GyongiG05} during the early days of web search, there was an increasing need to identify quality documents. Both document quality measures based on link analysis (such as PageRank) and those that rely on a document's content or metadata~\cite{DBLP:journals/ir/CormackSC11} are recognised as effective ways to identify documents crafted maliciously~\cite{DBLP:journals/computer/GyongiG05}. For instance, Bendersky et al.~\cite{10.1145/1935826.1935849}  examined features such as entropy, the proportion of stopwords in a document, and other lexical features within a learned document quality estimator (their estimator was used for ranking, but could also have been used for index pruning). Such content features were often used as components within learning-to-rank models~\cite{10.1145/2433396.2433407,ZhouDQ2005}, along with URL and link analysis features. Among many works on spam document detection, Cormack et al.~\cite{DBLP:journals/ir/CormackSC11} showed that a simple content- and metadata-based classifier could be used to prune many spam documents from a ClueWeb09 index.

Finally, while static pruning can be used for creating smaller indices, it also has a role in creating {\em tiered} index settings~\cite{Baeza-Yates:2008:DTS:1409220.1409223,ntoulas:2007,Rossi:2013:FDQ:2484028.2484085,Skobeltsyn:2008:RCR:1390334.1390359} - in such a setting, a smaller aggressively pruned index tier is used for answering common queries (perhaps located in faster storage media), while the larger index tier is retained for infrequent use by long-tail queries. As an illustration, Skobeltsyn et al.~\cite{Skobeltsyn:2008:RCR:1390334.1390359} demonstrated that employing a combination of term-centric and document-centric pruning in a tiered index setup enabled the management of 85\% of queries using only a quarter of the resources compared to the full index of the search engine.

Overall, no recent work has addressed the pruning of entire documents or passages from an index for modern retrieval engines. Our work aims to address this gap. Our work, however, is most concerned with the retrieval of focussed passages from documents.  Passages are often identified using a simple sliding window of tokens from documents~\cite{DBLP:conf/sigir/HearstP93,10.1145/3331184.3331303} or by leveraging the document's structure~\cite{DBLP:conf/sigir/Callan94}. A number of works have considered the identification of useful parts of documents. For instance, in a static pruning context, de Moura et al.~\cite{10.1145/1060745.1060783} aimed to identify {\em significant sentences} within a document for indexing---significant sentences are those containing the most important terms (as identified using term-centric pruning).

In terms of datasets, the well-known MSMARCO (v1) passage ranking dataset---which we use in our experiments---contains passages obtained for a large sample of queries from {\em "a state-of-the-art passage retrieval system at Bing"}~\cite{bajaj2016ms}; we believe that the passage extraction of this system includes some notion of passage quality (e.g., passages that are likely to contain answers to the queries from which the dataset is constructed), but the details of this system are not public. For this reason, we also include experiments addressing the generalisation of our approach on the MSMARCO v2 passage corpus~\cite{DBLP:conf/trec/Craswell0YCL21}, which was derived from a large set of seed documents, obtained using a {\em ``query-independent proprietary algorithm for identifying promising passages, and selected the [138M] best non-overlapping passages''}. With this in mind, it is clear that when evaluating our approach using MSMARCO passage datasets, these are actually settings where some low-quality passages have already been removed by a proprietary algorithm.

\section{Passage Quality Estimation}

\looseness -1 \crc{We now introduce our framework for passage quality estimation (Section \ref{sec:meth:prelim}), several baseline quality estimators (Sections \ref{ssec:lexical}--\ref{sec:meth:latent}), \crc{our proposed supervised quality estimator (Section \ref{sec:meth:sup})}, and our static pruning strategy (Section \ref{sec:meth:prune}).}

\subsection{Preliminaries}\label{sec:meth:prelim}

Let a corpus consist of a set of passages $P=\{p_1,p_2,...,p_n\}$ extracted from a set of documents (or a set of documents that a search engine will likely index). Further, consider a set of user queries $Q=\{q_1,q_2,...q_m\}$ that are issued (or likely to be issued) to a search engine. A traditional relevance model estimates the relevance of a specific passage $p_?\in P$ with respect to a specific query $q \in Q$. We use the notation $\textsc{Rel}(p_?\mid q)$ to represent this value, using nomenclature borrowed from probability (i.e., the relevance of $p_?$ conditioned on $q$).\footnote{We purposefully depart from the nomenclature of the Probability Ranking Principle~\cite{robertson1977probability} to both generalise to the vast array of modern relevance models (which often do not have a proper probabilistic interpretation) and to better highlight the centrality of the target passage $p_?$ to the relevance and quality estimations.} $\textsc{Rel}(p_?\mid q)$ implicitly produces a relevance score with respect to the full corpus of passages $P$, given that a subset of passages from $P$ are ultimately ranked against one another using this value.

\begin{table}
\centering
\caption{Examples of relevance and quality models by how they condition their measure of relevance. For example, DuoT5 directly conditions on both a specific query ($q \in Q$) and a reference passage ($p_1 \in P$): $\textsc{Rel}(p_?\mid q,p_1)$, while our QualT5 model conditions on neither a specific query nor specific reference passages: $\textsc{Qual}(p_?)$.}
\label{tab:cond}
\begin{tabular}{c|cc}
 & \multicolumn{2}{c}{\bf Query Condition} \\
\bf Passage & $\textsc{Rel}(\cdot)$ & $\textsc{Qual}(\cdot)$ \\
\bf Condition & $q\in Q$ & $\varnothing$ \\
\cmidrule(lr){2-3}
$\{p_1,p_2,...,p_n\}\subseteq P$ & RankGPT~\cite{DBLP:conf/emnlp/0001YMWRCYR23} & CDD~\cite{ZhouDQ2005} \\
$p_1\in P$ & DuoT5~\cite{DBLP:journals/corr/abs-2101-05667} & - \\
$\varnothing$ & MonoBERT~\cite{DBLP:journals/corr/abs-1901-04085} & ITN~\cite{DBLP:conf/sigir/ZhuG00}, QT5 (ours) \\
\end{tabular}
\end{table}

To produce a finer-grained relevance prediction, some relevance models further condition $\textsc{Rel}$ on one or more reference passages. A DuoEncoder~\cite{DBLP:journals/corr/abs-2101-05667} performs a relevance estimation of $p_?$ with respect to another passage $p_1$ in addition to $q$: $\textsc{Rel}(p_?\mid q,p_1)$. As a result, the predictions of a DuoEncoder are only meaningful with respect to other passages compared against the same query, and therefore either need to be aggregated across a consistent set of reference passages~\cite{DBLP:journals/corr/abs-2101-05667} or be applied in situations where a well-defined reference passage exists~\cite{DBLP:conf/sigir/MacAvaneyS23}. Recently, listwise scores further condition on a larger set of $n$ reference passages~\cite{DBLP:conf/emnlp/0001YMWRCYR23}: $\textsc{Rel}(p_?\mid q,p_1,p_2,...,p_n)$, providing further reference points, but also further constraining the settings under which the predictions are meaningful.

In contrast with this direction, a passage quality estimator $\textsc{Qual}(\cdot)$ \textit{reduces} the context upon which a relevance estimation is made by eliminating the direct condition on $q$. Instead, it implicitly produces an estimation of a passage's relevance across a population of all queries likely to be submitted to an engine $Q$, akin to how $\textsc{Rel}(p_?\mid q)$ implicitly compares against $P$. In its simplest form, $\textsc{Qual}(p_?)$ does not directly condition against any other text, though as will be shown in the following section, some forms of $\textsc{Qual}(\cdot)$ condition on statistics generated from the corpus $P$. Table~\ref{tab:cond} compares how different types of relevance and quality models directly condition on a query and/or other passages.

\subsection{Statistical Quality Estimators}\label{ssec:lexical}

Several approaches for estimating the quality of textual content based on lexical statistics \crc{already exist}. We explore two representative approaches as baselines in our work: the Information-To-Noise ratio (ITN)~\cite{DBLP:conf/sigir/ZhuG00} and the Collection-Document Distance (CDD)~\cite{ZhouDQ2005}. The Information-To-Noise (ITN) ratio is defined as the ratio of the number of unique terms in a passage to the total number of terms:
$ITN(p)=\frac{CountUniqueTerms(p)}{CountTotalTerms(p)}$.
Given that existing literature considers documents with a higher ITN to be of higher quality, we can use ITN directly as a passage quality function, i.e., $\textsc{Qual}_{ITN}(p_?):=ITN(p_?)$.

The Collection-Document Distance (CDD) takes a statistical language modeling perspective on the problem of document quality estimation. Specifically, it is defined as the KL-Divergence between the (smoothed) document language model and the collection language model:
$CDD(p_?\mid P)=\sum_{t\in T} Pr(t\mid P) \log\frac{Pr(t\mid P)}{\lambda Pr(t\mid p_?) + (1 - \lambda) Pr(t\mid P)}$,
where $T$ is the set of all tokens in the lexicon, $Pr(t\mid P)$ is the probability of token $t$ in the corpus $P$, $Pr(t\mid p_?)$ is the probability of token $t$ in the passage $p_?$, and $\lambda\in [0,1]$ is the language model's smoothing parameter.
Unlike ITN, CDD is directly conditioned on the passage corpus $P$. 
Since CDD measures the \textit{distance} between the corpus and a specific passage, we negate the CDD value when using it as a quality measure: $\textsc{Qual}_{CDD}(p_?):=-CDD(p_?)$, i.e., higher quality passages should be more similar to the corpus as a whole, rather than less similar.

\subsection{Unsupervised Neural Quality Estimators}

CDD relies on unigram ``bag-of-words'' language models instantiated using the target corpus itself. In contrast, learned neural language models are typically more useful, as they can measure the conditional likelihood of a sequence of words. To this end, we examine the use of a neural language model's perplexity as a measure of passage quality. Perplexity (denoted PPL) is measured as follows:
$PPL(p_?) = \text{exp} \left( -\frac{1}{\mid p_?|} \sum_{i}^{|p_?|} \log Pr\left(p_?[i]\mid p_?[<i]\right)  \right)$,
where $\log Pr(p_?[i]\mid p_?[<i])$ is the log-likelihood of token the $i$th token in the passage following the tokens $p_?[<i]$~\cite{chen1998evaluation}.

In essence, perplexity measures the inverse likelihood that the language model would generate the sequence of tokens. This is akin to CDD, essentially measuring the ``distance'' that a particular passage is from the data upon which the language model was trained. Therefore, we hypothesise that passages with high perplexity are of low quality, as they consist of unlikely or unusual sequences of tokens: $\textsc{Qual}_{PPL}(p_?):=-PPL(p_?)$. A passage quality estimator based on a pre-trained language model's perplexity has several advantages. First, it requires no supervision on relevance data, eliminating the data annotation and training costs associated with the latent and supervised models covered in the following sections. Further, the ubiquity of language models means that there are bespoke models available for a variety of domains~\cite{DBLP:conf/emnlp/BeltagyLC19}.

\subsection{Latent Neural Quality Estimators}\label{sec:meth:latent}
\looseness -1 Prior art suggests that trained neural relevance models already encode biases towards higher-quality passages~\cite{DBLP:journals/tacl/MacAvaneyFGDC22}. The next category of quality estimators attempt to extract this latent signal from relevance models that are already trained. Bi-encoder models, such as those that produce dense or learned sparse vectors, are reasonable candidates for extracting latent quality signals since all passage signals for a bi-encoder model are already encoded into that vector.

Assuming the typical inner product scoring method, scaling a passage vector's magnitude (while keeping its direction the same) will adjust its relevance score for all query vectors by that proportion. This quality makes a vector's magnitude a natural way for a model to latently encode a passage's query-independent quality signals. Therefore, we consider a bi-encoder's passage vector magnitude as a latent quality signal:
$\textsc{Qual}_{Mag}(p_?) := \left\Vert Encode(p_?) \right\Vert$,
where $Encode(p_?)$ returns a bi-encoder's passage vector.

On the other hand, to the best of our knowledge, the only bi-encoder model that explicitly encodes a latent document quality signal is EPIC~\cite{epic}. EPIC's architecture enforces a scale of its passage vectors by a value predicted in the range of [0,1]. Although this value is not supervised directly, we expect this scale to encode an overall passage quality score through its relevance training process, and we use it as an additional latent quality estimator.

\looseness -1 Latent neural quality estimators have desirable qualities. Most importantly, the latent signal may be trivially obtainable during indexing. For instance, vector magnitudes are a minor additional calculation to make atop a dense vector if it would have been calculated anyway. Similarly, EPIC's passage quality estimator can simply be captured when calculating the sparse representation during indexing. Further, if neural latent signals provide a strong quality signal, it would reduce the computational burden of training a bespoke supervised quality model---as the following section proposes.

\subsection{Supervised Neural Quality Estimators}\label{sec:meth:sup}

Finally, we consider an approach that produces passage quality scores through direct supervision. Consider a supervised model $M_\theta(p_?)\in\mathbb{R}$, which is parameterised by $\theta$ and produces a real-value response for input passage $p_?$. Through supervision over a set of training triples $(q,p_+,p_-)\in T$, where $p_+$ and $p_-$ are relevant and non-relevant passages to $q$, we optimise $\theta$ to estimate a query-independent quality score:
\begin{equation}
\underset{\theta}{\arg\min}\hspace{0.2em} \mathcal{L}(M_\theta(p_+),1) +
\mathcal{L}(M_\theta(p_-),0) \hspace{0.3em}\forall\hspace{0.3em} (q,p_+,p_-)\in T
\end{equation}
where $\mathcal{L}$ is a pointwise loss function (though the supervision process can be modified to use pairwise or listwise losses, too). Importantly, note that $M_\theta$ does not consider the query, only the passage. Once optimised, $M_\theta$ can be used directly as a quality estimator: $\textsc{qual}_{sup}(p_?):=M_\theta(p_?)$.

\subsection{Passage Quality for Static Pruning}\label{sec:meth:prune}

The primary application of a passage quality estimator is as a signal for pruning low-quality passages before indexing. Given a passage quality estimator $\textsc{Qual}(p_?)$ (or one conditioned on the entire passage corpus, like CDD, $\textsc{Qual}(p_?|P)$), a pruning strategy defines a quality threshold $t$ with which to sample the passage corpus:
\begin{equation}
P^\prime = \left\{
p \mid p \in P \land \textsc{Qual}(p) \geq t
\right\}
\end{equation}
Then, rather than indexing (and ultimately retrieving from) the entire corpus $P$, only the pruned subset $P^\prime$ is indexed. \crc{There are many options on how to choose the (model-dependent) quality threshold $t$: a search engine operator could choose this value empirically using sample documents of varying quality, choose it based on operational criteria (e.g., target index size), or choose it \textit{a priori} (given a well-calibrated model). In this work, we primarily aim to determine which quality methods provide the best tradeoffs overall, and therefore test each method at numerous values of $t$, each representing a target proportion of the corpus to prune. We leave strategies for estimating which value of $t$ will meet the particular requirements of an engine for future work.}

\section{Experimental Setup}

We conduct experiments to answer several research questions about our approach. We begin from an intrinsic perspective by asking:
\vspace{-0.25em}\begin{enumerate}
\item[\bf RQ1] Which of the presented passage quality estimators can best identify passages that do not answer any known query?
\end{enumerate}\vspace{-0.25em}
\looseness -1 Although intrinsic analysis gives an overview of the relative merits of the quality estimators, an extrinsic evaluation in the primary use case for these models---static pruning---is essential to understand their ultimate impact on search effectiveness. Hence, we ask:
\vspace{-0.25em}\begin{enumerate}
\item[\bf RQ2] When using passage quality estimators for static pruning, how much of a corpus can be pruned without affecting the precision of the top results?
\item[\bf RQ3] What are the computational overheads of applying quality estimators for passage pruning, and how do they affect a search engine's downstream computational and storage requirements?
\end{enumerate}\vspace{-0.25em}
Perhaps unsurprisingly, the supervised neural approach is far more effective than the other passage quality estimators. However, given that it also has a high inference cost, we ask:
\vspace{-0.25em}\begin{enumerate}
\item[\bf RQ4] What are the efficiency and effectiveness trade-offs of various model sizes for supervised passage estimators?
\end{enumerate}\vspace{-0.25em}
Finally, we test how well the best passage quality estimator transfers to corpora of other characteristics:
\vspace{-0.25em}\begin{enumerate}
\item[\bf RQ5] How well do supervised passage quality estimators transfer to corpora of other domains and/or scales?
\end{enumerate}\vspace{-0.25em}

\subsection{Datasets}\label{ssec:datasets}

We address RQ1-RQ4 using the well-understood MSMARCO (v1) retrieval dataset~\cite{bajaj2016ms}. We use the passages from the corpus extracted using their strong (proprietary) passage segmentation algorithm. We train our supervised QT5 estimator using the relevance labels from the MSMARCO training set. \crc{To understand the effects of quality pruning in both deep and shallow annotation settings,} we conduct evaluation using the TREC Deep Learning 2019~\cite{craswell2020overview} and 2020~\cite{Craswell2021OverviewOT} query sets (denoted DL 19 and DL 20, respectively), which have deep judgements (212.8 on average per query across the 97 queries), as well as the MSMARCO dev (small) query set, which has shallow judgements (1.1 across 6,980 queries). In our experiments, for reasons of brevity, we combine the results of the DL 19 and DL 20, as we found no notable differences between these query sets.

Noting both the unusual construction of the MSMARCO v1 corpus~\cite{bajaj2016ms} and the community's focus on the transferability of models across domains~\cite{DBLP:conf/nips/Thakur0RSG21}, we also test on the MSMARCO v2~\cite{DBLP:conf/trec/Craswell0YCL21} and CORD19~\cite{DBLP:journals/corr/abs-2004-10706} corpora. MSMARCO v2 allows us to demonstrate how well the approach scales to a more natural and substantially larger corpus (138M passages) while still building upon the high-quality industry passage segmentation applied by MSMARCO. We use the combined TREC DL 2021~\cite{DBLP:conf/trec/Craswell0YCL21} and 2022~\cite{DBLP:conf/trec/Craswell0YCLVS22} query sets (129 queries in total, with an average of 3,079 relevance assessments per query.\footnote{
Over the full, non-deduplicated corpus
}) To test the transferability to a new domain while maintaining relevance assessments with sufficient density, we use the full TREC COVID~\cite{DBLP:journals/sigir/VoorheesABDHLRS20} query set over CORD19 (50 queries with 1,386 assessments per query on average). We use the concatenation of CORD19 titles and abstracts, and truncate the documents to a model's maximum length---following a typical evaluation setting for this dataset. We use the natural language `description' queries. \crc{We note that although there are a multitude of IR test collections available~\cite{DBLP:conf/sigir/MacAvaneyYFDCG21}, these two are sufficient to assess generalisation RQ5.}

\subsection{Ranking Pipelines}\label{ssec:pipelines}

To validate the effect that passage quality pruning has on down-stream retrieval, we test the following retrieval pipelines, which are representative of several paradigms:

\noindent \textbf{ $\bullet$ BM25 (Lexical).} As a lexical retriever, we use BM25~\cite{DBLP:conf/trec/RobertsonWJHG94}. We build a BlockMaxWAND~\cite{DBLP:conf/sigir/DingS11} index using the PISA engine~\cite{pisa}.

\noindent \textbf{ $\bullet$ TAS-B (Dense).} As a dense retriever, we use TAS-B~\cite{tasb}.\footnote{\href{https://huggingface.co/sebastian-hofstaetter/distilbert-dot-tas_b-b256-msmarco}{\texttt{sebastian-hofstaetter/distilbert-dot-tas\_b-b256-msmarco}}\label{foot:tasb}} We perform an exact $k$-nearest neighbour search over a flat index to avoid variability introduced in using approximation techniques.

\noindent \textbf{ $\bullet$ SPLADEv2 (LSR).} As a learned sparse retriever, we use SPLADEv2 (lg)~\cite{DBLP:conf/sigir/LassanceC22},\footnote{\href{https://huggingface.co/naver/efficient-splade-VI-BT-large-doc}{\texttt{naver/efficient-splade-VI-BT-large-doc}} and \href{https://huggingface.co/naver/efficient-splade-VI-BT-large-query}{\ldots\texttt{-VI-BT-large-query}}} a model that balances efficiency, effectiveness, and representation size. We index and retrieve using PISA~\cite{pisa}.

\noindent \textbf{ $\bullet$ BM25 >> MonoELECTRA (Cross-Encoder).} Finally, we re-rank the top 100 BM25 results using MonoELECTRA~\cite{DBLP:conf/ecir/PradeepLZLYL22},\footnote{\href{https://huggingface.co/crystina-z/monoELECTRA_LCE_nneg31}{\texttt{crystina-z/monoELECTRA\_LCE\_nneg31}}} a strong cross-encoder model trained with hard negatives.

We conduct all our experiments using the PyTerrier~\cite{macdonald2020declarative} platform and we use language model implementations from the HuggingFace Transformers package~\cite{DBLP:journals/corr/abs-1910-03771}.

\subsection{Passage Quality Estimators}\label{ssec:estimators}

Throughout our experiments, we compare the following passage quality estimators:

\noindent \textbf{ $\bullet$ Random.} As a simple baseline, we uniformly randomly sample a passage's quality as a real number between 0 and 1. For consistency, we use the same random value for a passage across all experiments.

\noindent \textbf{ $\bullet$ Lexical (ITN and CDD).} As existing lexical baselines, we use ITN and CDD (described in Section~\ref{ssec:lexical}). We tokenise and use the Porter stemmer for both estimators. For CDD, we apply a language model smoothing of $\lambda=0.99$ after pilot experiments.

\noindent \textbf{ $\bullet$ Unsupervised (GPT2 and T5-Base Perplexity).} As existing unsupervised neural baselines, we use the perplexity of GPT2\footnote{\href{https://huggingface.co/gpt2}{\texttt{gpt2}}}~\cite{radford2019language} and T5-Base\footnote{\href{https://huggingface.co/t5-base}{\texttt{t5-base}}}~\cite{DBLP:journals/jmlr/RaffelSRLNMZLL20}, using the LM-PPL software package.\footnote{\href{https://github.com/asahi417/lmppl}{github.com/asahi417/lmppl}}

\noindent \textbf{ $\bullet$ Latent (TAS-B Magnitude and EPIC Quality).} As two latent neural indicators of passage quality, we use the magnitude of TAS-B~\cite{tasb} passage representations\footref{foot:tasb} and the passage quality weight from EPIC~\cite{epic}.\footnote{The EPIC model checkpoint provided by \citet{epic} was used.}

\noindent \textbf{$\bullet$ Supervised (QualT5).} Finally, as a supervised passage quality estimator, we fine-tune three sizes of a T5~\cite{DBLP:journals/jmlr/RaffelSRLNMZLL20,DBLP:journals/corr/abs-2109-10686} backbone model\footnote{\href{https://huggingface.co/t5-base}{\texttt{t5-base}}, \href{https://huggingface.co/t5-small}{\texttt{t5-small}}, \href{https://huggingface.co/google/t5-efficient-tiny}{\texttt{google/t5-efficient-tiny}}} over the MSMARCO's official training triples\footnote{\crc{We recognise that the negatives present in the triples are not necessarily low-quality. However, we expect this training setup to work nonetheless, given that relevance models trained using these potentially false negatives can also overcome this noise in the training data.}} (ignoring the query component). In line with models like MonoT5 and DuoT5~\cite{DBLP:journals/corr/abs-2101-05667}, we use a prompt structure of: \texttt{Document: [x] Relevant: [true/false]}. Due to observations showing that models can over-fit MSMARCO's training triples when training for too long~\cite{DBLP:conf/emnlp/NogueiraJPL20}, we train for 10k iterations with a batch size 16. \crc{We use cross entropy loss,} the Adam optimiser~\cite{DBLP:journals/corr/KingmaB14}, and a learning rate of $5\times 10^{-5}$.

\subsection{Measures}\label{ssec:measures}
Pruning quality is essentially a classification task: can we identify passages that will not be relevant to any queries. For an intrinsic passage quality evaluation, we consider all relevant passages known to us (the full dev set plus TREC DL 19 and 20\footnote{We exclude relevant passages found in the training set entirely, since several of our quality estimators use this data for training.}) and measure the classification accuracy of each passage quality estimator. In particular, we consider we plot the Receiver Operator Curve (ROC), and further report the Area Under the ROC Curve (AUC). %

\begin{figure}
\centering
\includegraphics[scale=0.63]{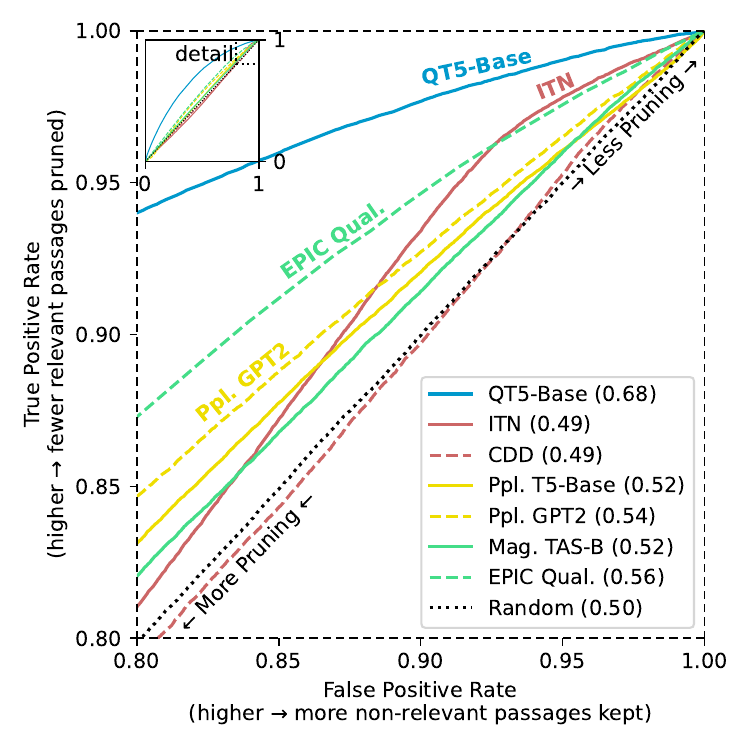}\vspace{-2em}
\caption{\looseness -1 ROC curves for each passage quality estimator, based on a union of all relevant documents in the full MSMARCO dev set,  DL 2019, and DL 2020 and excluding all relevant passages from the train set. The figure details the range [0.8,1.0], thereby focussing on the passages most likely to be pruned. The AUC for each estimator is in the legend.}
\label{fig:roc}
\end{figure}

\looseness -1 For our extrinsic evaluations, we measure ranking effectiveness on the four datasets previously mentioned: MSMARCO dev (small), TREC COVID, TREC DL 19+20, and TREC DL 21+22. We focus on the precision-oriented top-10 results using official task measures, which are often the most valuable for users and downstream AI agents~\cite{DBLP:conf/nips/LewisPPPKGKLYR020}. For MSMARCO dev (sm), we measure the mean reciprocal rank (\texttt{RR@10}), given the sparse assessments. For all other datasets, we use \texttt{nDCG@10}~\cite{DBLP:journals/tois/JarvelinK02}. Since in pruning settings we are primarily concerned with whether results are of the same quality as unpruned ones (instead of statistically different), we use a TOST equivalence test~\cite{schuirmann1987comparison} with $p<0.05$, a lower-bound threshold of 5\%, and an arbitrarily-large upper bound \crc{(since improved effectiveness is a desirable side-effect)} \crc{to establish statistical significance across RQs 2--5}. Although pruning can affect recall, in most question-answering retrieval settings like these, we are only concerned with recall insofar as it affects the precision of a later retrieval stage. Therefore, and considering the space it would take to present our results on recall, we consider our re-ranking pipeline sufficient for evaluating the effect of reduced recall.

\section{Results \& Analysis}

In this section, we answer our research questions and provide further analysis of the methods.

\subsection{RQ1: Intrinsic Effectiveness}

Figure~\ref{fig:roc} presents the ROC curve of each passage quality estimator and their corresponding AUC. We first observe that passage quality estimation is clearly a difficult task: most models only achieve a marginal improvement (if any) over the random baseline. The sole exception is the supervised QT5 model, with an AUC of 0.68. The latent EPIC Quality signal is the second strongest signal, with an AUC of 0.56, followed by unsupervised GPT2 perplexity at 0.54. Although ITN starts out as a very strong indicator (top right corner, second only to QT5 at a False Positive Rate (FPR) of 0.95), it quickly degrades to near-random effectiveness by a FPR of 0.8.

To answer RQ1, the supervised QT5 model provides a substantially stronger estimation of passage quality than the other methods. Nonetheless, other methods show reasonable potential---especially EPIC's quality signal---despite not being trained on the task directly.

\begin{figure*}
\centering
\includegraphics[scale=0.58]{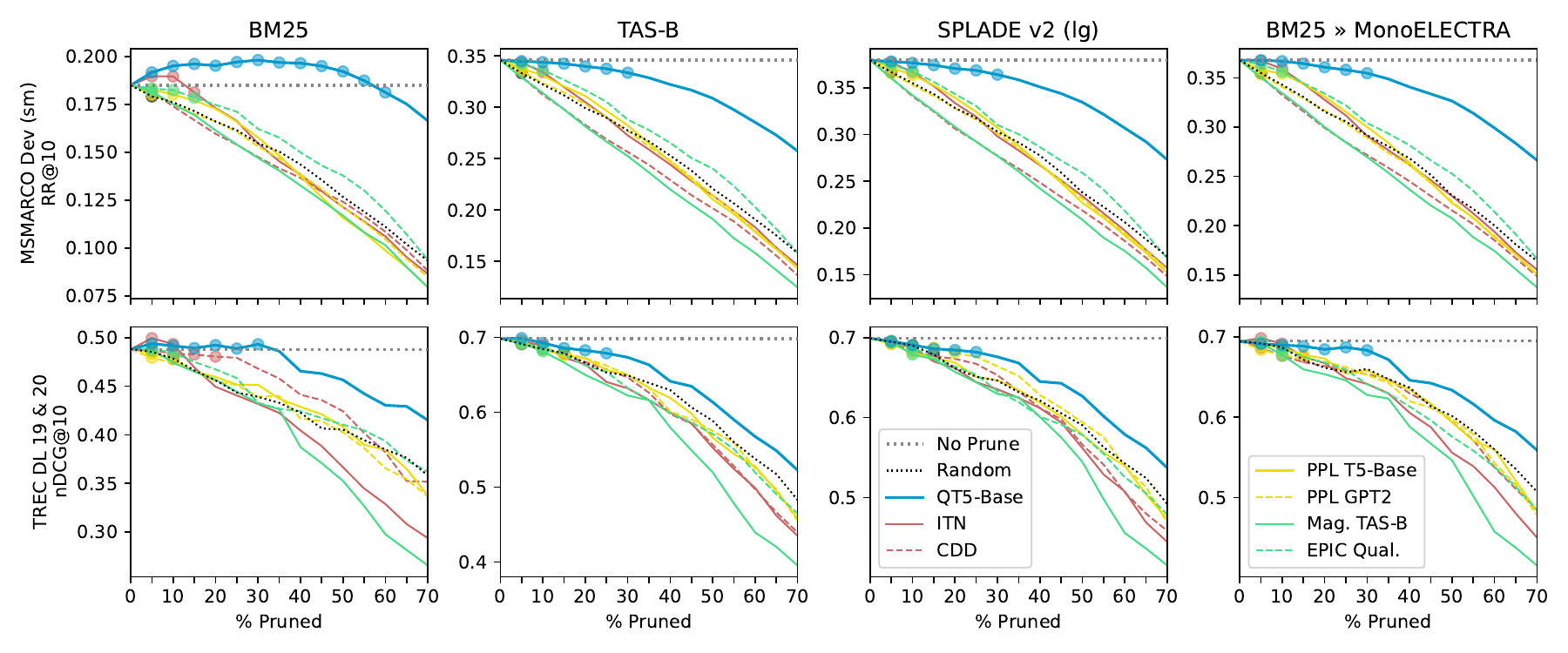}
\vspace{-2em}
\caption{Precision-oriented retrieval effectiveness on four pipelines by the percentage of a corpus pruned using each quality estimator. Effectiveness measurements that are statistically equivalent to the unpruned passage corpus are marked with $\medbullet$. \crc{Note that the vertical axis of each plot are scaled to emphasise the effect on each individual model.}}
\label{fig:approaches}
\end{figure*}

\subsection{RQ2: Pruning Effectiveness}

\begin{table}[b]
\centering
\caption{\looseness -1 The proportion of passages in MSMARCO that can be pruned while maintaining statistically equivalent effectiveness on MSMARCO Dev small (Dev) or DL 2019 \& 2020 (DL) using a given retrieval pipeline. Values in \red{grey} are less effective than random. The highest value in each column is \textbf{boldened}.}
\label{tab:how_much_pruning}
\vspace{-1em}
\resizebox{85mm}{!}{
\begin{tabular}{lrrrrrrrr}
\toprule
& \multicolumn{2}{c}{BM25} & \multicolumn{2}{c}{TAS-B} & \multicolumn{2}{c}{SPLADE} & \multicolumn{2}{c}{ELECTRA} \\
\cmidrule(lr){2-3}
\cmidrule(lr){4-5}
\cmidrule(lr){6-7}
\cmidrule(lr){8-9}
& Dev & DL & Dev & DL & Dev & DL & Dev & DL \\
\midrule
Random & 5\% & 10\% & 5\% & 15\% & 5\% & 15\% & 5\% & 10\% \\
\midrule
QT5-Base & \bf 60\% & \bf 30\% & \bf 30\% & \bf 25\% & \bf 30\% & \bf 25\% & \bf 30\% & \bf 30\% \\
ITN & 15\% & 10\% & 10\% & \red{10\%} & 10\% & \red{10\%} & 10\% & 15\% \\
CDD & 5\% & 20\% & \red{0\%} & \red{10\%} & \red{0\%} & \red{10\%} & \red{0\%} & 10\% \\
Ppl. T5-Base & 10\% & 10\% & 5\% & 15\% & 10\% & \red{10\%} & 10\% & 15\% \\
Ppl. GPT2 & 5\% & \red{5\%} & 5\% & 15\% & 5\% & 20\% & 5\% & 10\% \\
Mag. TAS-B & 5\% & \red{5\%} & \red{0\%} & \red{10\%} & \red{0\%} & \red{10\%} & \red{0\%} & 10\% \\
EPIC Qual. & 15\% & 10\% & 10\% & \red{10\%} & 10\% & \red{10\%} & 10\% & 10\% \\
\bottomrule
\end{tabular}}
\end{table}

Next, we move on to our primary extrinsic evaluation to test how valuable passage quality estimators are for static pruning. Figure~\ref{fig:approaches} shows the precision-oriented measures (RR@10 and nDCG@10) for each of our proposed passage quality estimators over four retrieval pipelines. Each estimator is used to prune the passages in the corpus down to at most 70\% of the original size, at steps of 5\%. By 70\%, all systems experience a significant degradation in effectiveness. However, several estimators can prune a substantial proportion of the corpus while maintaining significantly equivalent retrieval effectiveness. Most notably, QT5-Base can consistently prune 25\% or more of the corpus while maintaining equivalent effectiveness. \crc{For the most part, we observe similar behaviour between the sparsely judged (Dev) and deeply judged (DL) datasets.}

Table~\ref{tab:how_much_pruning} summarises the maximum proportion of the corpus that can be pruned while maintaining equivalent effectiveness. For instance, ITN can prune 15\% of the dev set when retrieving using BM25 while maintaining equivalent effectiveness. The table shows that QT5-Base is the only passage quality estimator that consistently outperforms the random baseline.

To answer RQ2, we find that QT5-Base can consistently prune 25-30\% of the MSMARCO passage corpus with statistically equivalent effectiveness, consistently outperforming the random baseline. Other estimators enable a lower proportion of a corpus to be pruned (e.g., EPIC Qual. allows 10-15\%), but do not consistently outperform the random baseline.

\subsection{RQ3: Pruning Efficiency}

Given that RQ2 shows that neural passage quality estimators can effectively prune a passage corpus without affecting retrieval effectiveness, we now investigate the computational overheads of applying them. Table~\ref{tab:eff} presents the average throughput (in passages per second) and corresponding mean latency (in milliseconds per passage) of each passage quality estimator.\footnote{Measurements were made by taking the highest throughput over 5 samples of 10,000 passages from MSMARCO on a commodity GPU (NVIDIA 3090) at 100\% (or nearly 100\%) utilisation. Furthermore, the batch size of each method was adjusted to maximise its throughput. ITN and CDD did not benefit from GPU hardware acceleration, so this estimator was done exclusively on CPU.}

We first observe that, despite the high effectiveness of QT-Base seen in RQ1 and RQ2, it adds considerable computational overhead at 1.46 ms/passage. As a point of reference, PISA indexes the MSMARCO passage dataset at 48,474p/s or 0.02ms/p on the same CPU (without needing a GPU). Therefore, in an effort to close the efficiency gap, we also explore two smaller versions of QT5 using smaller base models: QT5-Small and QT5-Tiny. These models offer substantially faster encoding times, reducing the mean latency down to 0.13ms/p for QT5-Tiny. Although this latency is still substantially higher than the cost of lexical indexing, it provides a competitive operating point for neural indexing approaches that also require model inference, such as dense or learned sparse approaches. For instance, the mean time to encode embeddings for the dense TAS-B model is 0.94ms/p.\footnote{The passage encoding cost dominates the total indexing costs for a flat dense index. Building approximate nearest neighbour index structures, such as HNSW~\cite{DBLP:journals/pami/MalkovY20} can add additional indexing time, which we ignore for this analysis.}

Although the cost of quality estimation is required for each passage in a corpus in a pruning setting, the more of the corpus that can be pruned, the lower the passage encoding time becomes since fewer passages need to be encoded. This means that quality estimators which cost less than an encoding model can break even---or even reduce---the total computational cost of indexing. We compute the \textit{break-even} point, which is the average percentage of a corpus that would need to be pruned to offset the added cost of quality estimation.\footnote{$BreakEven=\frac{QualityLatency}{EncodingLatency}$} Table~\ref{tab:eff} provides this ``break-even'' point for TAS-B and SPLADE v2. We observe that QT5-Tiny breaks even at 14\% for TAS-B and 13\% for SPLADE, with any additional pruning reducing the overall dense or learned sparse encoding time.

\begin{table}[b]
\caption{Throughput, mean latency, and break-even indexing points of each passage quality estimator. Rows with * indicate that the measurement was taken exclusively on CPU.
}
\label{tab:eff}
\vspace{-1em}
\centering
\begin{tabular}{lrrrr}
\toprule
\multicolumn{2}{r}{Throughput} & Latency & \multicolumn{2}{c}{Break-even \% Pruned} \\
\cmidrule(lr){2-2}
\cmidrule(lr){3-3}
\cmidrule(lr){4-5}
& p/s & ms/p & TAS-B & SPLADE (lg) \\
\midrule
QT5-Base & 683 & 1.46 & \red{>100\%} & \red{>100\%} \\
QT5-Small & 2,205 & 0.45 & 48\% & 45\% \\
QT5-Tiny & 7,716 & 0.13 & 14\% & 13\% \\
ITN* & 25,551 & 0.04 & 4\% & 4\% \\
CDD* & 8,616 & 0.12 & 13\% & 11\% \\
Ppl. GPT2 & 684 & 1.46 & \red{>100\%} & \red{>100\%} \\
Ppl. T5-Base & 581 & 1.72 & \red{>100\%} & \red{>100\%} \\
Mag. TAS-B & 1,064 & 0.94 & 100\% & 93\% \\
EPIC Qual. & 881 & 1.14 & \red{>100\%} & \red{>100\%} \\
\bottomrule
\end{tabular}
\end{table}

\begin{figure*}
\centering
\includegraphics[scale=0.55]{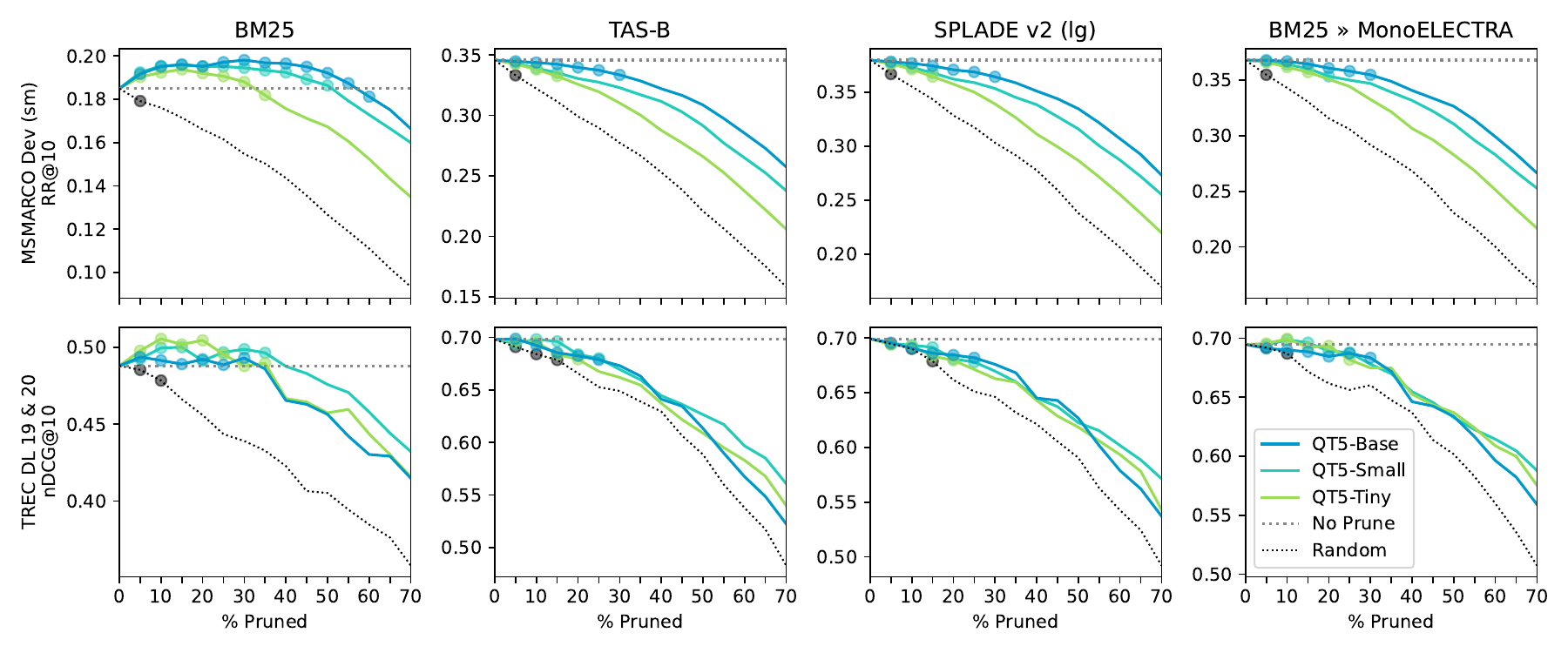}
\vspace{-1.6em}
\caption{Precision-oriented pruning effectiveness of three supervised QT5 model sizes on four pipelines. Effectiveness measurements that are statistically equivalent to the unpruned passage corpus are marked with $\medbullet$. \crc{Note that the vertical axis of each plot are scaled to emphasise the effect on each individual model.}}
\label{fig:sizes}
\end{figure*}

We now briefly explore how neural passage pruning can affect storage and retrieval overheads. Pruning inherently reduces the size of an index, since fewer passages are stored. For a flat single-representation dense index, this relationship is perfectly linear: each pruned passage reduces the size of the index by the same amount. However, the relationship is not necessarily linear for indexing techniques with variable document sizes, such as lexical or learned sparse indexes, since the pruning methods may (for instance) favour longer passages. Nonetheless, we find that the total index size deviates less than 5\% from a linear expectation when pruning at 70\% or below for both lexical and SPLADE indexes.\footnote{When pruning to over 70\%, the size of the lexicon begins to have a substantial impact on the total index size for MSMARCO.} Finally, the relationship between an index size and its retrieval costs are well-understood~\cite{DBLP:conf/sigir/MacdonaldTO12,DBLP:journals/pami/MalkovY20}: as the size of an index decreases (e.g., number of postings in a sparse index or number of vectors in a dense one), so does retrieval time. Concretely, we observe that pruning a PISA BM25 index with QT5-Tiny at 25\% reduces the index size by 24.3\% and reduces retrieval time by 9.8\%.

In summary, to answer RQ3, passage pruning strategies can reduce an engine's storage costs and retrieval times. Meanwhile, they can also \textit{reduce} the indexing time for engines that need to encode documents using a neural network---e.g., dense or learned sparse models---as long as the quality model is sufficiently small (e.g., QT5-Tiny). Next, we will explore the effect that reducing the size of the supervised model has on pruning effectiveness.

\begin{figure*}
\centering
\includegraphics[scale=0.58]{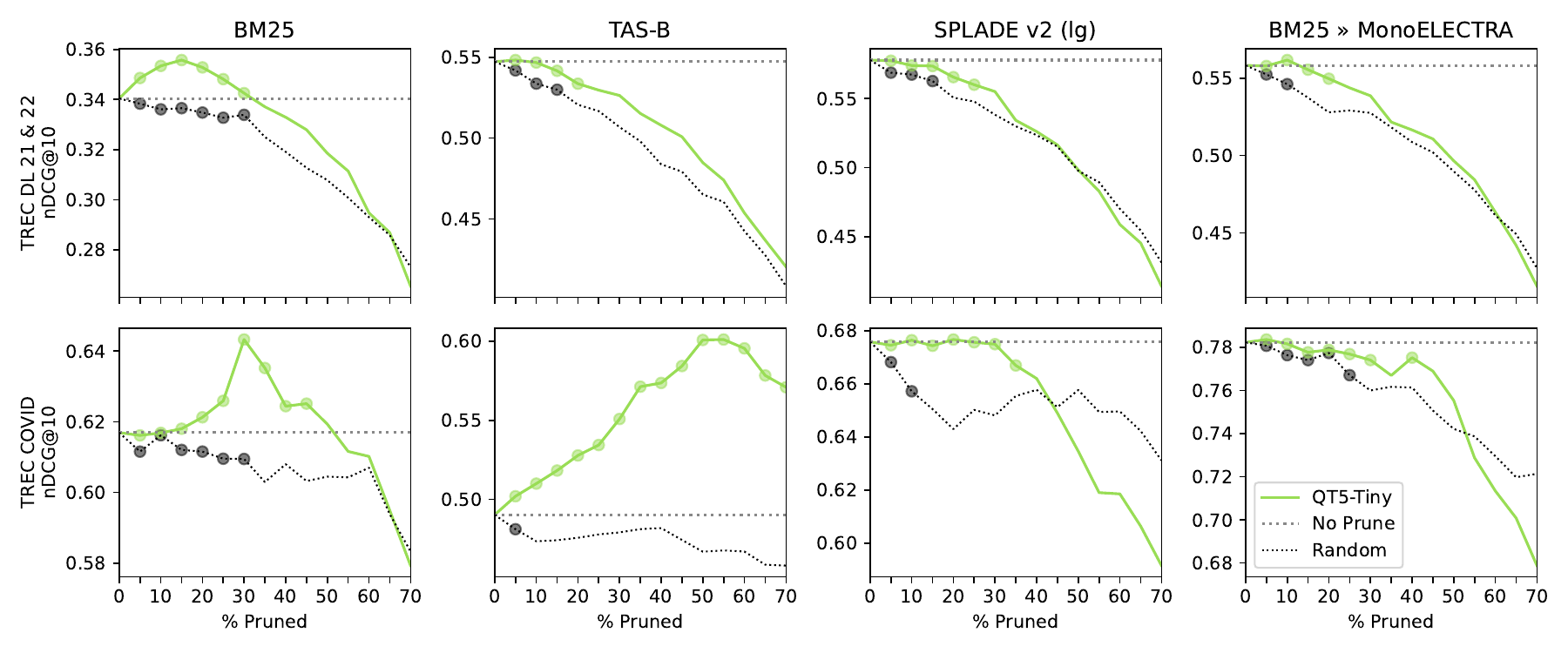}
\vspace{-1.5em}
\caption{Transferability of QT-5-Tiny to two other datasets: MSMARCO v2 (TREC DL 21\&22) and CORD19 (TREC COVID). Effectiveness measurements that are statistically equivalent effectiveness to the unpruned corpus are marked with $\medbullet$. \crc{Note that the vertical axis of each plot are scaled to emphasise the effect on each individual model.}}
\label{fig:transfer}
\end{figure*}

\subsection{RQ4: Supervised Model Sizes}

Given that RQ3 found that smaller supervised model sizes could yield substantial benefits in terms of efficient indexing of a corpus, we now explore the effects that smaller supervised models have on \crc{(extrinsic\footnote{\crc{The \href{https://github.com/terrierteam/pyterrier-quality/blob/main/figures/roc.all.pdf}{online appendix} provides an intrinsic analysis that draws the same conclusions.}})} pruning effectiveness. Figure~\ref{fig:sizes} presents the results for the base, small, and tiny QT5 models on Dev and DL 19\&20. On Dev, we see a consistent trend: the larger the model, the more of a corpus it can successfully prune while maintaining equivalent effectiveness. For instance, on a BM25 pipeline, the base, small, and tiny models can prune 60\%, 50\%, and  35\%, respectively. However, this trend largely disappears when evaluating on TREC DL 19\&20, which have more complete relevance assessments. For instance, on a BM25 pipeline, the small and tiny models can prune to 35\%, while the base model only can prune to 30\%. These results suggest that the tiny model is just as capable of estimating passage quality, for the most part, as the base model, and the differences observed on Dev may just amount to the base model's higher capacity to over-fit to the training data distribution (the Train and Dev sets are both sparsely annotated and sampled from the same distribution).

\looseness -1 To answer RQ4, we find that smaller (i.e., more computationally efficient) supervised quality models are mostly just as capable as larger ones at estimating passage quality, while larger models may be more susceptible to over-fitting. Indeed, the QT5-tiny model, which can estimate passage quality in 9\% of the time of the QT5-base model, can create a smaller index than when using the base model without exhibiting significantly degraded effectiveness compared to the unpruned index, according to the TREC DL 19\&20 dataset.

\subsection{RQ5: Transferability}

Up to this point, we have conducted experiments using version 1 of the MSMARCO passage dataset. However, we recognise the oddities in construction that this dataset has. Therefore, we test whether QT5-Tiny\footnote{We focus on this passage quality estimator because RQ3 established its efficiency and RQ4 established its effectiveness. Meanwhile, the other quality estimators either were unable to outperform a random baseline consistently (ITN, CDD, Ppl, Mag. TAS-B, and EPIC Qual.) or were more expensive than QT5-Tiny without providing consistently stronger effectiveness (QT5-Base and QT5-Small).} can transfer to a substantially larger corpus constructed using a more conventional method (MSMARCO v2 using TREC DL 21\&22) and one from a different domain (CORD19 using TREC COVID). Figure~\ref{fig:transfer} presents the pruning results of QT5-tiny compared to a random pruning baseline. On TREC DL 21\&22, we see that QT5-Tiny can consistently prune 20\% or more of the corpus while maintaining equivalent effectiveness, similar to the observations made on TREC DL 19\&20. Meanwhile, on TREC COVID, we see that methods that are initially weak (BM25 and TAS-B) can actually substantially improve the search result quality by pruning. This observation is especially apparent for TAS-B, which has been seen to previously over-fit to MSMARCO~\cite{DBLP:conf/nips/Thakur0RSG21}. Meanwhile, the stronger SPLADE and MonoELECTRA pipelines can prune 30\% or more of the corpus while maintaining equivalent effectiveness.

\looseness -1 To answer RQ5, we find that QT5-Tiny can successfully transfer to a larger corpus and a corpus to a different domain.

\subsection{Analysis: Examples of Pruned Passages}

We now briefly explore some examples of passages with low quality estimations to gain additional insights into what is lost through the pruning process. QT5-Tiny's lowest-quality passage from MSMARCO v1 is passage \texttt{16467}: \textit{``Back symptoms and Graves disease and Joint pain and Abnormal blood test symptoms (2 causes)...''}. This is an excerpt from an index page that routes users to the causes of various symptoms related to back pain, but does not provide any information directly. The first passage with a known positive label is \texttt{7195070}: \textit{``HBO HD Channels 899: HBO HD 900: HBO HD West 902: HBO 2 HD 903: HBO 2 HD...''}, which is the model's 1287th lowest-quality passage. It is part of a channel directory listing, and does, in fact, answer query \texttt{1100076} \textit{``hbo hd channel number''}. We observe that the passages with the lowest quality estimations tend to include lots of repetition. Other examples of low-scoring passages that probably ought not to be pruned include those with song lyrics that are particularly repetitive. The passage with the 5\% lowest quality is \texttt{5488158}: \textit{``Live and learn. Give two weeks notice, but not more. ^^ Good point. Would it be recommended to tell your next employer when they ask a good start date, I've going to give them two weeks, but they are probably going to ask me to leave that day, can we be flexible in the start date.''} This passage doesn't contain the repetition seen in the lowest-quality passages, but still does not provide much information that would likely be a useful search result to most question-style queries. Similarly, the passage with the 20\% lowest quality is \texttt{5543358}: {\em ``Don’t do it to lose weight or because you feel pressured to join a gym; rather, do it because it’s something you enjoy doing...''} which does not provide much of an answer on its own without additional context. Overall, this qualitative analysis suggests that the QT5-Tiny model identifies passages that the authors consider low-quality. However, there is also room to improve passage quality models to better handle edge cases, such as where passages are repetitive but still provide helpful information.

\section{Conclusion}

In this work, we demonstrated that neural methods---especially those supervised on relevance signals---can provide strong query-independent passage quality estimations. These quality signals can be used for substantial passage-level pruning of a corpus, saving indexing, storage, and retrieval costs while maintaining statistically equivalent effectiveness \crc{across a variety of corpora and retrieval pipelines.} \crc{These savings are important because they ultimately impact the power consumption and carbon footprint~\cite{DBLP:conf/sigir/ScellsZZ22} of operating AI-powered search engines.}

We recognise several limitations of this work, which are good candidates for future investigation. For instance, we primarily focus on settings that already have strong passage segmentation performed. We note that our approach has potential to be even more effective in settings where such segmentation is less reliable. \crc{Another possible direction is to incorporate these quality signals into the segmentation process itself, with the aim of splitting documents into the most meaningful passages and disregarding the rest.} Further, our supervised method is relatively na\"ive, simply training on relevance labels. Future work can investigate more advanced training techniques, such as distillation from the target ranker, a common practice when training relevance models~\cite{lin2021batch}. Finally, although we already demonstrate that our static pruning method works with a dynamic pruning retrieval algorithm (BMW~\cite{DBLP:conf/sigir/DingS11}), there is still room to test the compositionality of our passage-based pruning approach with other pruning paradigms, such as term-centric pruning~\cite{10.5555/1763653.1763665}, \crc{including in dense settings~\cite{DBLP:conf/doceng/AcquaviaMT23} or learned sparse~\cite{DBLP:conf/sigir/MackenzieDGC20,DBLP:conf/sigir/LassanceLDCT23} settings.}

\crc{Nevertheless, given the substantial savings we observed using relatively simple methods, we hope that this work sets the stage for future research in information retrieval that focuses more on what content from a corpus is worth indexing, rather than how to best retrieve from a fixed corpus.}

\bibliographystyle{ACM-Reference-Format}
\bibliography{biblio}

\end{document}